\documentclass[12pt,a4paper]{article}
\usepackage{epsfig}
\usepackage{epic}
\usepackage{float}
\usepackage{afterpage}
\usepackage{latexsym}
\restylefloat{figure}
\renewcommand{\baselinestretch}{1.7}\normalsize

\newcommand{\be}{\begin{equation}}
\newcommand{\ee}{\end{equation}}
\newcommand{\bea}{\begin{eqnarray}}
\newcommand{\eea}{\end{eqnarray}}
\newcommand{\klgl}{\:\hbox to -0.2pt{\lower2.5pt\hbox{$\sim$}\hss}
{\raise3pt\hbox{$<$}}\:}
\newcommand{\grgl}{\:\hbox to -0.2pt{\lower2.5pt\hbox{$\sim$}\hss}
{\raise3pt\hbox{$>$}}\:}
\newcommand{\kzwei}[1]{K_2\left( #1 \right)}
\newcommand{\gtap}{\raisebox{-.4ex}{\rlap{$\sim$}} \raisebox{.4ex}{$>$}}

\begin{document}
%
%
\markboth{ }{ }
\renewcommand{\baselinestretch}{1}\normalsize
\vspace*{-2cm}
\hfill TUM-HEP-357/99

\vspace*{2cm}
\bigskip
\bigskip
\begin{center}
{\huge\bf{Dynamics of Metastable Vacua
in the Early Universe}} 
\end{center}
\bigskip
\begin{center}
{\large{Bastian Bergerhoff}}\footnote{
bberger@physik.tu-muenchen.de}, 
{\large{Manfred Lindner}}\footnote{
lindner@physik.tu-muenchen.de} 
{\large{ and Manfred Weiser}}\footnote{
mweiser@physik.tu-muenchen.de}
\vspace*{0.3cm}\\
Physik Department, Technische Universit\"at M\"unchen \\
James-Franck-Strasse, D-85748 Garching, Germany
\end{center}
\setcounter{footnote}{0}
\bigskip
\vspace*{1cm}\begin{abstract}
\noindent
We study the question whether a possible metastable vacuum state 
is actually populated in a phase transition in the 
early universe, as is usually assumed in the discussion of vacuum 
stability bounds e.g. for Standard Model parameters. A 
phenomenological $(3+1)$-dimensional Langevin equation is solved
numerically for a toy model with a potential motivated by the 
finite temperature 1-loop effective potential of the Standard Model
including additional non-renormalizable operators from an effective 
theory for physics beyond the Standard Model and a time dependent 
temperature. It turns out that whether the metastable vacuum is 
populated depends critically on the value of the phenomenological 
parameter $\eta$ for small scalar couplings. For large enough 
scalar couplings and with our specific form of the  non-renormalizable 
operators the system (governed by the Langevin equation) always 
ends up in the metastable minimum.
\end{abstract}
\vspace*{1cm}
PACS-No.s: 11.10.Wx, 64.60.-i, 64.60.My
\newpage
\renewcommand{\baselinestretch}{1.1}\normalsize

{\bf{1.}} 
The symmetry breaking potential of a field theory at vanishing 
temperature $T$ may have local (energetically disfavoured) minima 
in addition to the global minimum which defines the true vacuum. 
This leads to metastability if the system sits at $T=0$ in a false 
vacuum and the decay of such a false vacuum may occur in two ways.
First it is possible to cross the classical energy barrier which 
results from the balance between the released volume energy of a 
single ``bubble of the correct vacuum'' with the amount of energy 
stored in the surface of that bubble. Vacuum decay sets in after
formation of the first bubble of critical size which requires 
however a big enough thermal fluctuation or some source of local 
energy deposition. Alternatively the false vacuum may decay via 
quantum tunneling and the parameters of the theory determine in
general which decay mechanism dominates.

The study of questions concerning the decay of metastable vacua under 
the influence of fluctuations induced by quantum effects goes back 
to the work of Krive and Linde \cite{KL} who studied vacuum stability 
under quantum corrections in the $\sigma$-model. The issue of 
vacuum stability may also be relevant for the Standard Model 
(see e.g.~the review of Sher \cite{Sher} and references therein) 
and for model extensions. Requiring that our electro-weak vacuum 
is the true vacuum leads for a large top mass $m_t \gtap 80$~GeV
to lower bounds on the Higgs mass \cite{LSZ}. These bounds can be 
somewhat relaxed by allowing the electro-weak vacuum to be metastable 
with a lifetime longer than the age of the universe \cite{Arnold},
where the dominating decay mode is the classical transition via 
the formation of a critical bubble triggered by random collisions 
of ultra high energy cosmic rays. Due to the experimental lower bounds on the 
Higgs mass metastability is in the Standard Model almost ruled out.
It may however still be an interesting issue in different model 
extensions and is certainly also in general an interesting question
in field theory.

Thermal effects were included only in a few cases in studies of 
metastability (see e.g. \cite{Arnold:1991cv}-\cite{Nikos1}). 
Most authors 
concentrated on transitions at vanishing temperature assuming that
the universe is already in the false vacuum. This ignores however
that the symmetry was restored at high temperature in the early 
universe and we will discuss in this paper the question how likely
or unlikely it is that the system falls into a metastable vacuum
with cosmological boundary conditions. Given the physical  
situation realized in the early universe (or more generally in a 
system that is cooled through a phase transition), we calculate 
thus for a theory exhibiting symmetry breaking with metastable 
vacua at $T=0$ the probability to end up in the false vacuum 
in the first place.

To address this topic one needs to study the response of a given 
field theory to a changing temperature which is obviously a highly 
non-trivial task\footnote{Since the temperature in the situation considered
here explicitely depends on time, one can not take recourse to well established
approaches like Langers theory of bubble nucleation. Langers description may
however be used in the late stages of the evolution when the time-dependence of
the temperature is dynamically no longer important and if the potential barrier
seperating the vacua is sufficiently high such that a saddle point expansion
about the top of the barrier is a good approximation
\cite{NAC}.}. A calculation from first principles would require
to self-consistently identify a heat bath, to calculate the 
coupling between the bath and the system and solve the equations 
of motion for the macroscopic expectation values that describe the 
system as the temperature of the heat bath is changed. All this 
would have to be done in the framework of non-equilibrium field
theory. The conceptual foundations for such a study are in principle 
given (at least for a scalar theory) by Morikawa \cite{Mor} and 
Gleiser and Ramos \cite{GR} (c.f.~also Boyanovsky et.~al.~ \cite{Boy}).
Using the CTP-formulation of field theory by Schwinger and Keldysh 
\cite{SK} they discuss the (perturbative) formulation of an effective 
equation of motion in a real $\lambda\varphi^4$-theory which turns 
out to resemble the well known Langevin equation in the form
\be
\Box \varphi + V'(\varphi) + \eta(\varphi) \dot{\varphi} = \xi(\varphi)~,
\label{Ltype}
\ee
albeit with rather complicated dissipative and noise contributions.

This makes contact with a number of numerical studies that use
equations of the type given in eq.~(\ref{Ltype}) to analyze the decay 
of false vacua at finite temperature \cite{57}-\cite{121}. In \cite{23} 
and \cite{121}, for example, (3+1)-dimensional theories are studied 
using an equation of the form 
\be
\Box \varphi + U'(\varphi,T) + \eta \dot{\varphi} = \xi~,
\label{Lsimple}
\ee
where the used effective potential $U(\varphi,T)$ is motivated by the
high temperature result found in the Standard Model written as
\be
U(\varphi,T) = \frac{a}{2} ( T^2-T_2^2 ) \varphi^2 - \frac{\alpha}{3} T
\varphi^3 + \frac{\lambda}{4} \varphi^4~,
\label{UhighT}
\ee
where $\eta$ and $\xi$ are related by the fluctuation -- dissipation relation 
\be
\langle \xi(\vec{x},t) \xi(\vec{x}',t') \rangle = 2 \eta T \delta^{(4)} (x-x')
\qquad .
\label{fdr}
\ee
The authors studied the system at a fixed (namely the critical) 
temperature and choose $\eta$ to be $1$, arguing that the value of 
$\eta$ would not affect the final (thermodynamical) state of the 
system, but only the approach to equilibrium.
Here we will use a Langevin equation of the form (\ref{Lsimple}) 
to model the dynamics of a scalar theory with a symmetry breaking 
phase transition (at some critical temperature $T_c$) and with 
metastable vacua for $0 \leq T (\grgl T_c)$ in order to calculate 
the probability of metastable states. As mentioned above the 
question is whether the metastable states decay during cooling 
by thermal activation long before the temperature approaches zero 
or if the system populates the metastable states with reasonable
probability which would be the starting of the usual vacuum 
stability discussion in the Standard Model.

{\bf{2.}}
We first discuss the choice for the effective potential
$U(\varphi,T)$ appearing in equation (\ref{Lsimple}) and 
for the time dependence of the temperature. Even though 
we will not aim at a discussion of the full Standard Model, 
embedded in some specific extension at a given scale 
$\Lambda$ which renders the physical vacuum metastable, 
it makes sense to stick as close as possible to the
Standard Model potential. For reasons to be discussed below, 
we will only be interested in theories with a relatively 
small scalar self-coupling. We chose therefore as a first 
contribution to $U(\varphi,T)$ the 1-loop effective
potential of the Standard Model in the form (c.f.\ \cite{12})
\bea
U_{(1)}(\varphi,T) &=& - D T_2^2 \varphi^2 + \frac{\lambda}{4} \varphi^4 + B
\varphi^4 \ln \frac{\varphi^2}{\sigma^2} + \nonumber \\
&& +  \sum_{\mathrm{Bosons}} \frac{g_B T^4}{2 \pi^2} \int_0^\infty dx x^2 \ln
\left( 1 - e^{-\sqrt{x^2+(m_B/T)^2}} \right) - \nonumber \\
&& - \sum_{\mathrm{Fermions}} \frac{g_F T^4}{2 \pi^2} \int_0^\infty dx x^2 \ln
\left( 1 + e^{-\sqrt{x^2+(m_F/T)^2}} \right) 
\label{U1}
\eea
where $g_{B,F}$ count the number of degrees of freedom of the corresponding
fields and, taking into account only the $SU(2)$ gauge bosons and the
top quark, 
\bea
D &=& \frac{1}{24} \left[ 6 \left( \frac{m_W}{\sigma} \right)^2 + 3 \left(
\frac{m_Z}{\sigma} \right)^2 + 6 \left( \frac{m_t}{\sigma} \right)^2 \right]
\nonumber \\
B &=& \frac{1}{64 \pi^2 \sigma^4} \left( 6 m_W^4 + 3 m_Z^4 - 12 m_t^4 \right)
\nonumber \\
T_2^2 &=& \frac{m_H^2 - 8 B \sigma^2}{4 D} 
\label{parsU1}
\eea
with the vacuum expectation value $\sigma = 246$~GeV.

The influence of new physics from an extended model at some scale
$\Lambda$ can be parametrized by suitable higher dimensional 
non-renormalizable operators. For a specific extension these 
operators can be obtained as usual by connecting tree graphs which 
contain light scalars by propagators of heavy degrees of freedom.
For illustration one can assume heavy scalars with 
masses $M^2 \sim \Lambda^2 + T^2$ which interact with the light
scalars. To obtain the leading contribution to the
six point function one connects two quartic vertices which contain 
one heavy scalar. For small external momenta this results at $T=0$
in an effective six point function for the light scalar of the 
form $\varphi^6/\Lambda^2$. Including temperature effects results 
approximately in $\varphi^6/(\Lambda^2+T^2)$ and for small temperatures
we can finally expand in $T/\Lambda$.

For $\varphi \ll \Lambda$ the discussion can be restricted to
the leading higher dimensional operators. While contributions to the
two and four point functions can be absorbed into the definition of 
the usual scalar parameters the leading new terms occur in
the $6$- and $8$-point functions. This procedure yields thus an 
additional contribution to the potential of the form
\bea
U_{(2)}(\varphi,T) &=& \left(1-\frac{T^2}{\Lambda^2} \right) \left[ -
\frac{g_6}{6} \frac{\varphi^6}{\Lambda^2} + \frac{g_8}{8}
\frac{\varphi^8}{\Lambda^4} \right] 
\label{U2}
\eea
which is also heuristically suggestive. For the present study  
$g_6$ and $g_8$ should be viewed as couplings which parametrize
new physics. The common $T$-dependent
factor has the following properties: it vanishes for $T \to \Lambda$
where the theory is not reliable any more and it becomes unity for $T
\to 0$ as in the usual zero temperature field theory.

Finally, the interesting temperatures are of the order of the weak scale,
$T \sim {\mathcal{O}}(100)$~GeV. Taking the time dependence of the 
temperature to be governed by the (adiabatic) expansion of the universe 
according to the Friedman equations (curvature effects can be neglected 
at these temperatures  to a very good approximation) we have for a 
radiation dominated universe with thermodynamical equation of state
$\rho = 3 p$ that
\be
T \propto t^{-1/2} \qquad .
\label{Toft}
\ee
Using Gaussian white noise obeying the fluctuation -- dissipation relation
(\ref{fdr}) the only free parameters are now the scalar self-coupling $\lambda$
(resp.~the Higgs mass $m_H$),  the couplings $g_6$, $g_8$ and the scale
$\Lambda$, as well as the viscosity coefficient $\eta$. We consider 
different values of $\lambda$, taking $\lambda$ between $\lambda = 0.03$ 
(since for the physical Higgs mass one has 
$m_H^2 = (2 \lambda + 12 B) \sigma^2$ with $B$ as in eq.~(\ref{parsU1}) 
which is $B \simeq -0.0045$ if the measured masses are used, $\lambda$ 
has to be larger than $0.028$ for $m_H$ to be real) and -- for reasons 
to be discussed below -- $\lambda = 0.044$, corresponding to $m_H$ 
between $18.4$ and $45.1$~GeV.  For every value of $\lambda$, we choose 
values of $g_6$ and $g_8$ such that the full effective potential 
$U(\varphi,T) = U_{(1)}(\varphi,T) + U_{(2)}(\varphi,T)$ for $T=0$ 
has in addition to the physical minimum at $\sigma$ a second, deeper 
one at $\sigma'$ close to the scale of new physics $\Lambda$, which we 
choose to be at $\Lambda = 4 \sigma$.  This situation is reminiscent of 
the scenarios considered in connection with the stability bounds as 
discussed above. Finally $\eta$ is varied for each $\lambda$ in order 
to identify possible different dynamical regimes. 

{\bf{3.}}
Before discussing 
the results, let us give some details of the numerical procedure used. 
The full details may be found in \cite{Manfred}.
We want to perform numerical simulations in our simplified model inspired by
the Standard Model in an early universe context. Therefore the system is
rewritten in dimensionless quantities and we adopt the choice of \cite{23} 
by writing explicitly
\bea
  \tilde{x} := a^{1/2} T_2 x \qquad &,& \qquad \tilde{t}:= a^{1/2} T_2 t \nonumber \\
  X \equiv \tilde{\varphi} := a^{- 1/4} T_2^{-1} \varphi \qquad &,& \qquad
  \theta \equiv \tilde{T}  :=  T/T_2 \nonumber \\
  \tilde{\alpha} := a^{-3/4} \alpha \qquad &,& \qquad \tilde{\lambda}_{(T)} :=
  a^{-1/2} \lambda_{(T)} \nonumber \\
  \tilde{g_6} := a^{-1/2} g_6 \qquad &,& \qquad \tilde{g_8}  :=  a^{-1/2} g_8
\label{dimless}
\eea
where $a = 2D$ and 
\bea
\alpha = \frac{1}{4\pi} \left[ 6 \left(\frac{m_W}{\sigma}\right)^3+
                               3 \left(\frac{m_Z}{\sigma}\right)^3\right]
                               \qquad .
\label{alpha}
\eea
Omitting the twiddles, in the 
dimensionless system eq.~(\ref{Lsimple}) has the form
\be
\label{Lsimple2}
\frac{\partial^2 X}{\partial t^2} - \nabla^2 X + \eta \frac{\partial
X}{\partial t} + \frac{\partial U(X, \theta)}{\partial X} = \xi({\bf x}, t)
\ee
with the fluctuation -- dissipation relation
\be
\left< \xi({\bf x}, t) \xi({\bf x'}, t') \right> = 2 \eta \theta \delta(t -
t') \delta^{(3)}({\bf x - x'}) \qquad .
\ee

Instead of evaluating the full expression for the potential (\ref{U1}) at
any temperature we will for simplicity rely on the high and low temperature
expansion (HT and LT) \cite{12} in the appropriate regions of $m/T$. 
It is well known that both the high and the low temperature expansion 
work surprisingly well even for $m/T$ of order 1. Taking into account 
a field independent contribution to the HT-potential (which is often 
ignored) HT and LT can therefore be matched in a reasonably smooth 
manner at $m=T$. Even though there are different masses entering the 
potential we may exploit the fact that all masses are proportional to 
the scalar expectation value and use the top mass to decide upon 
which of the two regimes is appropriate. The potential of 
eq.~(\ref{U1}) can now be written as
\begin{eqnarray}
  \tilde{U}_{(1)}(X, \theta=0) & = & - \frac{1}{2} X^2 + \frac{1}{4}
  \tilde{\lambda} X^4 + \frac{B X^4}{\sqrt{a}} \ln \left(
  \left(\tilde{\lambda} + \frac{2 B}{\sqrt{a}} \right) X^2 \right) \nonumber \\
  \tilde{U}_{(1),{\mathrm{HT}}}(X, \theta) & = & \frac{1}{2} \theta X^2 - \frac{1}{3}
  \tilde{\alpha} X^3 + \frac{1}{4} (\tilde{\lambda}_T - \tilde{\lambda}) X^4
  - \frac{13}{60 a^{3/2}} \pi^2 \theta^4 \nonumber \\
  \tilde{U}_{(1),{\mathrm{LT}}}(X, \theta) & = & - \frac{\theta^4}{\pi^2 a^{3/2}} \left[
  6 \left( \frac{m_t}{\sigma} a^{1/4} \frac{X}{\theta} \right)^2 \cdot
  \kzwei{\frac{m_t}{\sigma} a^{1/4} \frac{X}{\theta}} \right. \nonumber \\
  & & \left. {} + 3 \left( \frac{m_W}{\sigma} a^{1/4} \frac{X}{\theta}
  \right)^2 \cdot \kzwei{\frac{m_W}{\sigma} a^{1/4} \frac{X}{\theta}}
  \right. \nonumber \\
  & & \left. {} + \frac{3}{2} \left( \frac{m_Z}{\sigma} a^{1/4}
  \frac{X}{\theta} \right)^2 \cdot \kzwei{\frac{m_Z}{\sigma} a^{1/4}
  \frac{X}{\theta}} \right] \nonumber  \\
  \tilde{U}_{(2)}(X, \theta) & = & \left( 1 - \frac{\theta^2}{\tilde{\Lambda}^2
  \sqrt{a}} \right) \left[ - \frac{1}{6} \tilde{g}_6
  \frac{X^6}{\tilde{\Lambda}^2} + \frac{1}{8} \tilde{g}_8
  \frac{X^8}{\tilde{\Lambda}^4}  \right] 
\end{eqnarray}
where $K_{\nu}(x)$ are the modified Bessel functions of the second kind
and of the order $\nu$. Since we need later on the first and second 
derivatives of $U$, it is more appropriate to use $K_2$ explicitely instead
of the more familiar exponential representation. The choices for $g_6$ and 
$g_8$ depending on $\lambda$ will be given below. Note the explicit time 
dependence of the potential appearing in eq.~(\ref{Lsimple2}) which is a 
consequence of the $t$-dependence of the temperature as given by 
eq.~(\ref{Toft}). 

This equation is analytically not solvable and we use therefore 
a discretization scheme for a numerical integration. Consider 
a hyper-cubic lattice in 3+1 dimensions with $N_x$ points in each 
spatial direction separated by the spacing $\delta x$ and $N_t$ points 
in the temporal direction separated by the spacing $\delta t$. 
Then $X_{i,n}$ is the value of $X$ at the lattice point 
$({\bf x_i}, n)$ where $i = (i_x, i_y, i_z)$. For sufficiently small
spacings the derivatives can be written as differences whose exact form 
depends on the discretization scheme. Like in \cite{23}, we use the well 
known staggered leapfrog scheme which is of second order accuracy in time.
Next the second order differential equation~(\ref{Lsimple2}) is rewritten 
as a pair of two first order equations for $X$ and $\dot{X}$. Introducing 
additional lattices for $\dot{X}$ which are shifted by $\delta t/2$ (i.e. 
$\dot{X}_{i,n+1/2}$ and $\dot{X}_{i,n-1/2}$) and applying the substitution 
\hbox{$X_{i,n+1} = X_{i,n} + \delta t \cdot \dot{X}_{i,n+1/2}$} yields
\begin{eqnarray}
  \label{eq:kap_43_6}
  \dot{X}_{i,n+1/2} & = & \frac{1}{1 + \frac{1}{2} \eta \delta t} \left[
    \left( 1 - \frac{1}{2} \eta \delta t \right) \dot{X}_{i,n-1/2}
    {} + \delta t \left( \nabla^2 X_{i,n} - \left. \frac{\partial U}{\partial
    X}\right|_{i,n} + \xi_{i,n} \right) \right] \quad . \nonumber \\
  & & 
\end{eqnarray}
The noise term becomes in discretized form 
\be
  \xi_{i,n} = \sqrt{\frac{2 \eta \theta}{\delta t (\delta x)^3}}
  \mathcal{G}_{i,n}
\ee
and thus the fluctuation -- dissipation theorem now reads
\be
\left< \xi_{i_1,n_1} \xi_{i_2,n_2} \right> = 2 \eta \theta \frac{1}{\delta t}
\delta_{n_1,n_2} \frac{1}{(\delta x)^3} \delta_{i_1,i_2} ~,
\ee
where the $\mathcal{G}_{i,n}$ are Gaussian distributed random numbers of 
unit variance. The initial conditions are chosen to be
\be
X_{i,0} = 0 \quad \mbox{and} \quad \dot{X}_{i,-1/2} = 0 ~,
\ee
which corresponds to a homogeneous system without fluctuations supposedly 
describing a rapidly and adiabatically cooled (i.e. quenched) state of 
the early universe before the onset of the electro-weak phase transition.
We also impose periodic boundary conditions to the finite lattice,
i.e. compactification of the spatial dimensions. The generation of unwanted
long range correlations is avoided due to the uncorrelated noise terms.

The simulation requires a few more words of explanation. Since the 
temperature evolves with time we have to fix the initial and the final 
temperature, $\theta_{{\mathrm{ini}}}$ and $\theta_{{\mathrm{fin}}}$, 
respectively, which are connected by
\begin{equation}
  \theta_{{\mathrm{ini}}} \cdot t_{{\mathrm{ini}}}^{1/2} = \theta_{{\mathrm{fin}}} \cdot t_{{\mathrm{fin}}}^{1/2}
\end{equation}
yielding thus the number $N_t$ of time steps $\delta t$,
\begin{equation}
N_t = \frac{t_{{\mathrm{fin}}} - t_{{\mathrm{ini}}}}{\delta t} 
  = \left[\left(
  \frac{\theta_{{\mathrm{ini}}}}{\theta_{{\mathrm{fin}}}}\right)^{1/\kappa} 
  - 1 \right] \cdot
  \frac{t_{{\mathrm{ini}}}}{\delta t} \quad .
\end{equation}
We choose for all calculations $\theta_{{\mathrm{ini}}} = 1.5$, well above 
the onset of the phase transition. The final temperature is to a large 
extent free; one just has to take care that all interesting dynamics has 
happened already when it is reached. $\theta_{{\mathrm{fin}}}$ should 
preferentially not be too small since $N_t$ depends strongly on the 
ratio $\theta_{{\mathrm{ini}}}/\theta_{{\mathrm{fin}}}$. Usually, 
$\theta_{{\mathrm{fin}}} = 0.15$ is a good compromise, together with 
$\delta t = 0.1$. The temperature $\theta(t)$ and thus the potential 
$U(X, \theta(t))$ with its first and second derivative are calculated
at every time step of the simulation. All extrema of $U$ are searched 
and it is determined whether they are minima or maxima. This fixes the 
number of phases which may be one, two or three in our model. Note that 
for $\theta > 0$ the second (or third) minimum may be at field values 
which are larger than the cutoff $\Lambda$. 

{\bf{4.}}
In order to discuss the results of the simulations we need to specify 
first values for the parameters of the potential. As described above, 
we assume the non-renormalizable operators introduced in eq.~(\ref{U2}) 
to be generated by new physics with characteristic scale
$\Lambda = 4 \sigma \simeq 1$~TeV. The corresponding couplings will 
be chosen such that for any given value of $\lambda$ the absolute 
minimum of the full potential is at $\sigma' \approx \Lambda$ for 
vanishing temperature. Finally $\lambda$ is varied as mentioned above 
between $0.03$ and $0.044$. Table 1 shows the values of the couplings 
$\lambda$, $\tilde{g}_6$ and $\tilde{g}_8$ together with the 
corresponding values of $m_H$ and the position of the stable minimum 
at $T=0$, $\sigma'$. The corresponding potentials at vanishing
temperature for the parameters of table 1 are plotted in figure 1.
Note that height of the barrier between the metastable minimum at 
$\sigma = 246$~GeV and the stable minimum at $\sigma'$ increases 
with growing $\lambda$.
\begin{table}
\begin{center}
\begin{tabular}{|c|c|c||c|c||c|}
\hline
   $\lambda$ & $\tilde{g}_6$ & $\tilde{g}_8$ & $m_H$ (GeV) & $\sigma'$ (GeV) &
   $\eta_{\mathrm{crit}}$ \\
\hline
    0.030 & 0.003941 & 0.063056 & 18.4 & 948.2 & 0.69 \\
    0.031 & 0.003892 & 0.062279 & 21.4 & 941.6 & 0.60 \\
    0.032 & 0.003578 & 0.057254 & 24.1 & 958.2 & 0.54 \\
    0.033 & 0.003558 & 0.056929 & 26.4 & 948.1 & 0.45 \\
    0.034 & 0.003549 & 0.056791 & 28.7 & 938.1 & 0.40 \\
    0.035 & 0.003285 & 0.052557 & 30.7 & 951.2 & 0.34 \\
    0.036 & 0.003290 & 0.052645 & 32.6 & 937.4 & 0.27 \\
    0.037 & 0.003054 & 0.048859 & 34.4 & 949.2 & 0.21 \\
    0.038 & 0.003066 & 0.049053 & 36.1 & 933.0 & 0.15 \\
    0.039 & 0.002806 & 0.044899 & 37.8 & 949.6 & 0.12 \\
    0.040 & 0.002818 & 0.045082 & 39.3 & 930.6 & 0.10 \\
    0.041 & 0.002620 & 0.041916 & 40.9 & 940.6 & 0.08 \\
    0.042 & 0.002436 & 0.038978 & 42.3 & 948.8 & 0.06 \\
    0.043 & 0.002400 & 0.038404 & 43.7 & 933.3 & 0.01 \\
    0.044 & 0.002227 & 0.035640 & 45.1 & 941.3 &  --  \\
\hline
\end{tabular}
\caption{\small Potential parameters (c.f.~eq.s (\ref{U1})-(\ref{U2}) and
(\ref{dimless})) and corresponding critical values of $\eta_{\mathrm{crit}}$}
\end{center}
\end{table}
\begin{figure}[ht]
  \begin{center}
    \includegraphics[width=10cm]{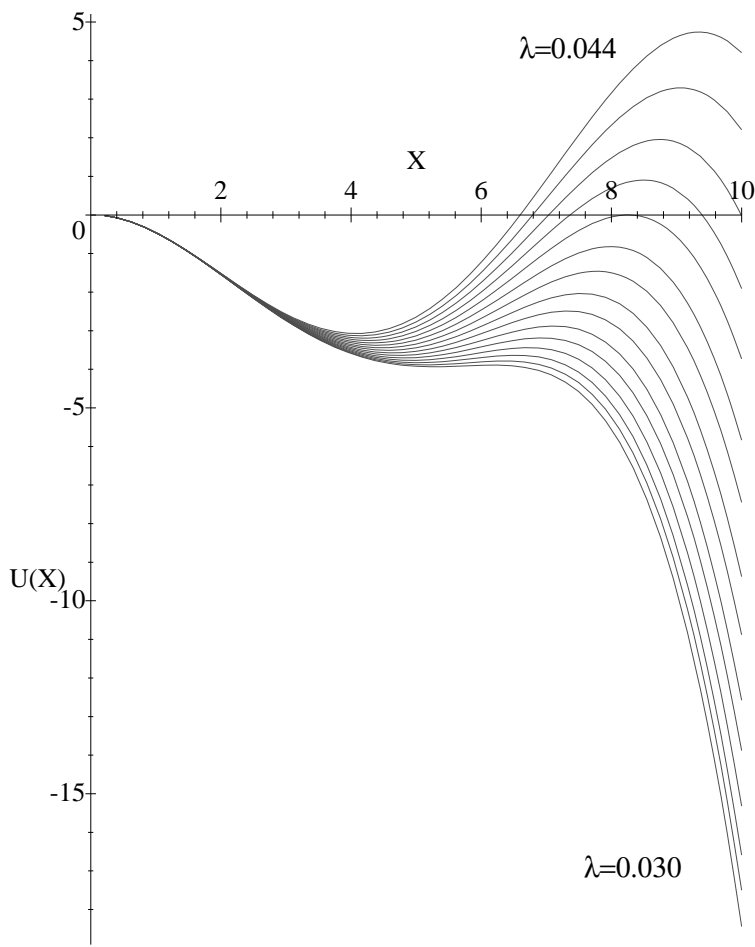}
    \caption{\small The dimensionless potentials $U$ as functions of the 
    dimensionless field $X$ for vanishing temperature with $\tilde{g}_{6,8}$ 
    as given in table 1. With increasing values of $\lambda$ the height of 
    the barrier between the metastable and the true vacuum gets larger.} 
  \end{center}
\end{figure}
The system is prepared with $X = \dot{X} = 0$ at $\Delta t = 0$, and the
temperature is taken to be $1.5 \times T_2$ initially. 
The temperature is then taken to depend on time according to eq.~(\ref{Toft})
and the Langevin equation is solved as described above.
A typical result of a solution of equation (\ref{Lsimple}) with the above
assumptions on the potential and for different values of $\eta$ is displayed 
in figure 2, where we plot the first and second moment of the dimensionless 
field, namely the mean value of the field $\langle X \rangle$ and the second 
moment $\Delta X = \sqrt{\langle (X - \langle X \rangle )^2 \rangle}$ as 
functions of $t$ for $\lambda = 0.03$ and different values of $\eta$. 
After an initial phase where the behaviour of the system is rather independent
of $\eta$ and where it evolves essentially homogeneously ($\Delta X$ is very 
small), the behaviour of the first moment shows a strong dependence on the
damping parameter $\eta$: For small $\eta$ the system reaches the absolute
minimum relatively fast and in a more or less homogeneous fashion, performing
damped oscillations around $X=\tilde{\sigma}'$.
As $\eta$ gets larger, the system remains close to the metastable minimum at
$\tilde{\sigma} \approx \tilde{\sigma}'/4$ for longer times and the transition
to the absolute minimum, though still occuring for $\eta < 0.7$, takes longer
and longer (note how the second moment deviates from $0$ for longer times --
the system is highly inhomogeneous during the transition). 
For $\eta = 0.7$, the behaviour changes dramatically: The system is no longer
able to cross the potential barrier and to populate the true vacuum, but 
remains in the metastable minimum with only very small fluctuations. 
The region of $\eta$ around the critical value $\eta_{\mathrm{crit}} = 0.69$ is
resolved in the lower panel of figure 2. 
\begin{figure}
  \begin{center}
    {\includegraphics[width=10cm,angle=-90]{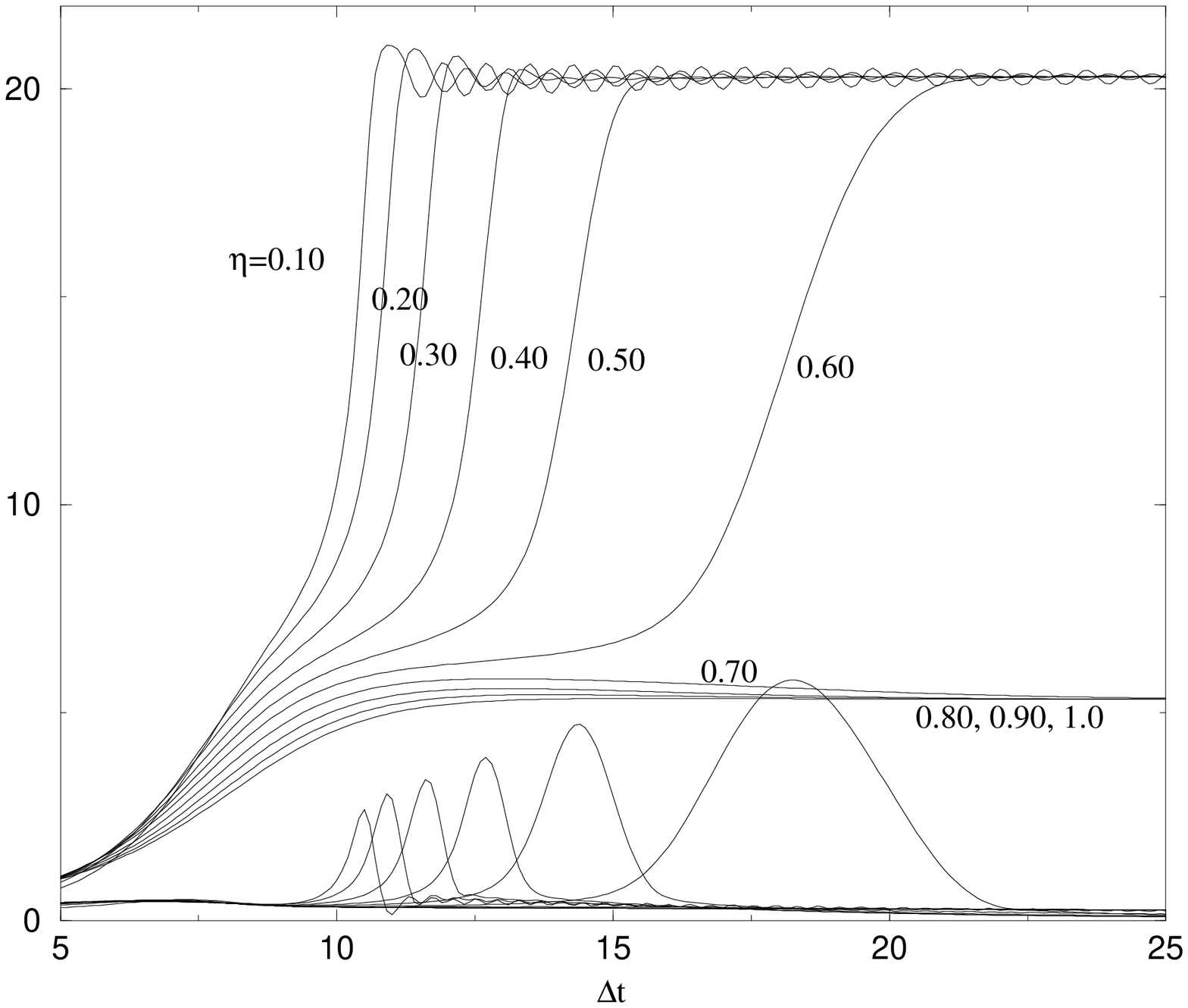}}
    {\includegraphics[width=10cm,angle=-90]{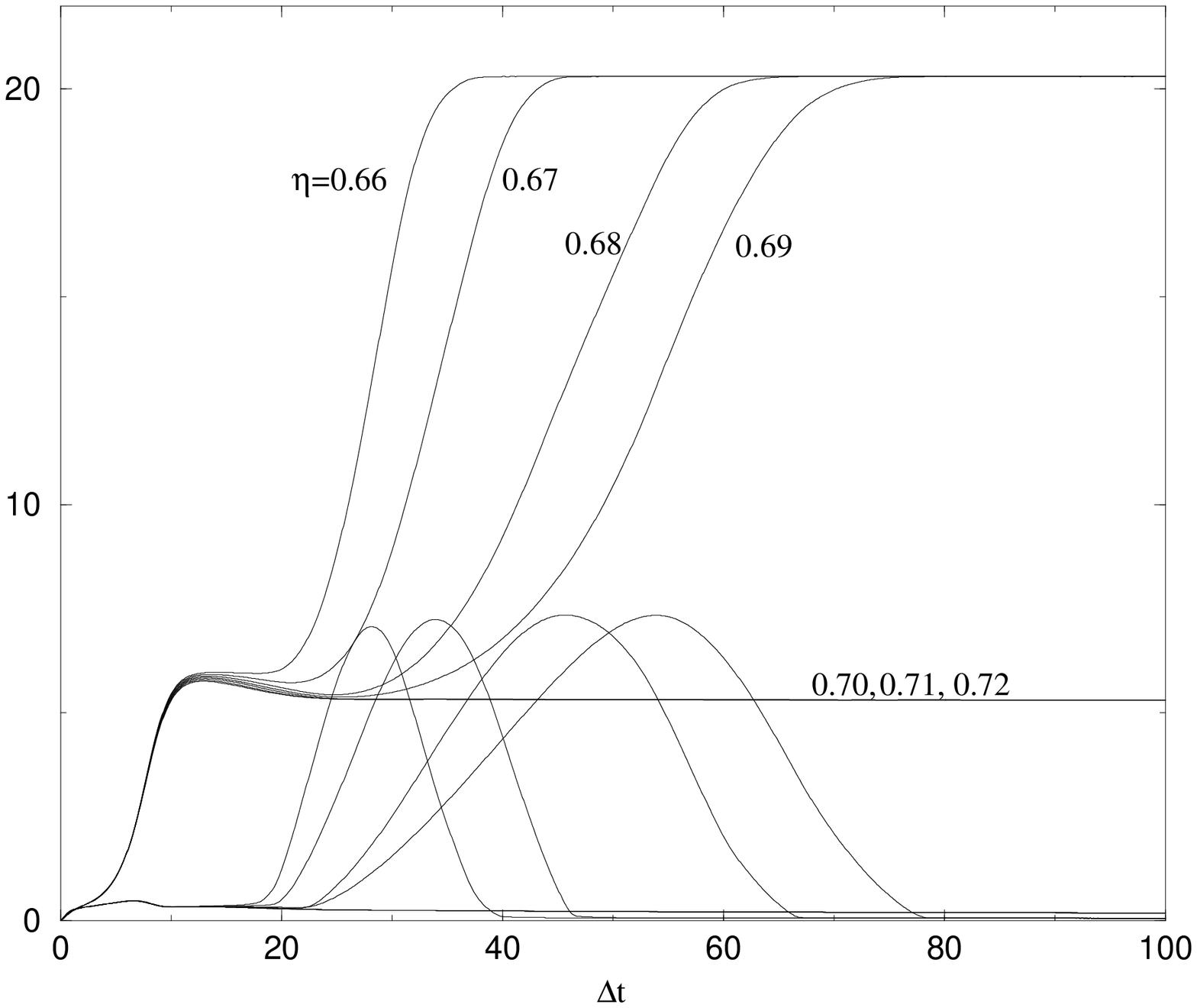}}
    \caption{\small The first and second moments $\langle X \rangle$ (upper 
    curves) and $\Delta X$ (lower curves) for $\lambda = 0.03$. In the 
    lower panel the dependence on $\eta$ around $\eta_{\mathrm{crit}} = 0.69$
    is resolved (note the different timescales).}
  \end{center}
\end{figure}
We perform the same procedure for the sets of parameters as given in the table,
always identifying the critical value of $\eta$ as long as it exists.
From table 1, one notices that $\eta_{\mathrm{crit}}$ is a decreasing
function of $\lambda$. In figure 3 we display the first moment for a 
relatively large coupling $\lambda = 0.42$. The critical $\eta$ turns out 
to be $0.06$ in this case. The dynamics of the system differs from the 
case displayed in figure 2 mostly by more pronounced oscillations around 
the false vacuum state for small $\Delta t$. These may be understood by 
noting that first the values of $\eta$ displayed are small compared to the 
ones shown in fig.~2, i.e.~the system is less damped, and that second the 
curvature of the potential at the metastable minimum is larger for larger 
$\lambda$ (c.f.~figure 1). By a simple kinematic analogon one expects more 
pronounced oscillations.
\begin{figure}
  \begin{center}
    \includegraphics[width=10cm,angle=-90]{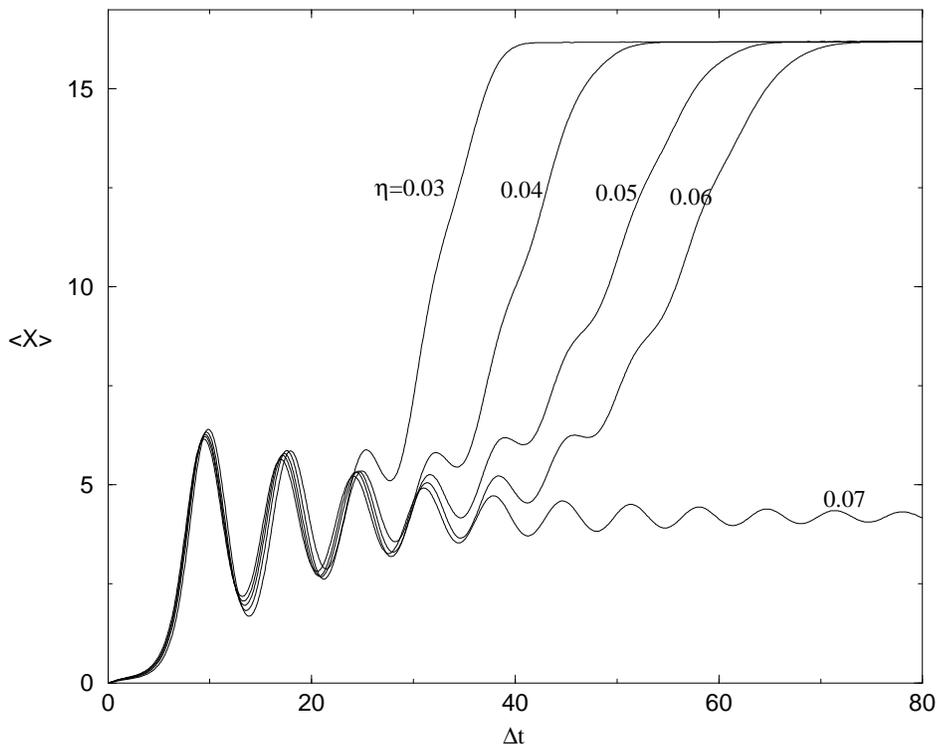}
    \caption{\small The first moment for $\lambda = 0.042$.}
  \end{center}
\end{figure}
Finally, for $\lambda = 0.044$, the critical value of $\eta$ vanishes and 
the system is no longer able to reach the true vacuum by thermal activation,
irrespective of the value of $\eta$.

We have checked finite size effects and found them to be small for the
lattice size chosen. In particular the values for $\eta_{\mathrm{crit}}$ 
found on a lattice with $64^3$ spatial sites are identical up to $\pm 0.01$ 
with the ones found on a $50^3$ lattice and given in the table. We have 
also performed calculations with time steps smaller than $0.1$ as used
for the results given above. Thus e.g.~the results for the moments at 
$\lambda = 0.03$ and $\eta$ around $0.69$, where one expects the errors
of discretization to be most pronounced, are identical within the 
numerical precision for $\delta t = 0.1$, $0.05$, and $0.03$. 
Too large values for $\delta t$ result of course in large deviations 
from the behaviour found above. We thus believe that the numerical error 
on $\eta_{\mathrm{crit}}$ (for the potential used) is $\pm 0.01$ 
and thus well under control.

{\bf{5.}}
To conclude, we have discussed the dynamics of the population and decay 
of metastable vacua in an environment with a time dependent temperature.
We have done this using a phenomenological Langevin equation for a scalar 
field with an effective potential motivated by the Standard Model embedded 
in some model with characteristic scale $\Lambda \sim 1$~TeV, leading 
for vanishing temperature to the existence of a stable vacuum state at 
$\varphi = \sigma' \sim \Lambda$.  This is the situation assumed for the
derivation of bounds on Standard Model parameters from vacuum stability 
as considered in the literature \cite{Sher}. We have discussed the 
dependence of the dynamics on the (phenomenological) parameter $\eta$, 
governing dissipative effects. In particular to answer to 
the question whether the 
theory, starting at high temperatures in the symmetric phase, populates 
the (now metastable) physical vacuum depends critically on this parameter. 
In the framework studied here, using an equation of motion of the form
(\ref{Lsimple}) with Gaussian white noise $\xi$ related to $\eta$ by
(\ref{fdr}) and a potential given by (\ref{U1})-(\ref{U2}) we find that the
metastable vacuum is indeed populated for any value of $\eta$ if $\lambda \grgl
0.043$, corresponding to $m_H \grgl 43.7$~GeV.
We have shown that for smaller values of $\lambda$ there exists a critical
value of the viscosity parameter, $\eta_{\mathrm{crit}}$, such that for $\eta <
\eta_{\mathrm{crit}}$ the system reaches the stable vacuum by thermal
activation before $T=0$, whereas for $\eta > \eta_{\mathrm{crit}}$ the system
will populate the metastable vacuum (being closer to $\varphi=0$ in our case)
and is not able to cross the potential barrier by thermal activation.

Clearly the results collected in table 1 depend both on the specific form of
the Langevin equation and on the effective potential. 
However, they highlight the fact that one may under certain conditions hope to
learn something about the dynamics of phase separation without detailed
knowledge of the parameters of the effective equation of motion.
Thus, while for small $\lambda$ the late time behaviour of the model studied
here crucially depends on whether $\eta$ is larger or smaller than
$\eta_{\mathrm{crit}}$, for large enough quartic coupling the value of $\eta$
is indeed irrelevant.
For a given theory it might thus suffice to calculate the effective potential
and its dependence on the temperature if the parameters of the theory are in a
range where $\eta_{\mathrm{crit}}=0$.
With the specific assumptions made in this work our results thus justify the
usual discussion of metastability in the Standard Model.
On the other hand we also note that for potentials with multiple minima the
final state of the system -- neglecting quantum effects, i.e.~in a treatment on
the basis of Langevin type equations -- is not generally independent of $\eta$
if the system is "quenched", i.e.~the temperature is time dependent.
This fact should motivate more efforts along the lines of 
\cite{Mor}, \cite{GR} 
and
\cite{Boy}
in order to improve our understanding of the derivation of phenomenological
equations of motion and their parameters from field theory.

\bigskip

{\bf{Acknowledgment:}} We would like to thank Nikos Tetradis for helpful
discussions.
This work was supported by the 
"Sonderforschungsbereich 375-95 f\"ur Astro-Teilchenphysik" 
der Deutschen Forschungsgemeinschaft.

\end{document}